\documentclass[prb,aps,epsfig,twocolumn,showpacs]{revtex4-1}
\usepackage{color}
\usepackage{cancel}
\usepackage{lipsum}
\usepackage{epsfig,amssymb,amsmath,latexsym}
\graphicspath{{chhobi/}}
\usepackage{float}
\usepackage{color}
\usepackage[colorlinks=true,linktoc=page,linkcolor=magenta,citecolor=magenta]{hyperref}
\begin{document}
\textheight=23.8cm

\title{Dynamics of impurity in the environment of Dirac fermions }
\author{ Ajit Kumar Sorout$^{*\dagger}$, Surajit Sarkar$^{\dagger}$ and Suhas Gangadharaiah$^{\dagger}$}
\affiliation{$^*$Department of Physics, University of Massachusetts, Amherst, MA 01003, USA}
\affiliation {$^\dagger$Department of Physics, Indian Institute of Science Education and Research, Bhopal, India}

\date{\today}
\pacs{1...}

\begin{abstract}  

We study the dynamics of a non-magnetic impurity interacting with the
surface states of a 3D and 2D topological insulator (TI).  
Employing the  linked cluster technique we develop a formalism for obtaining the Green's function of the mobile impurity interacting with the low-energy Dirac fermions.  
We show that for the non-recoil case in 2D, similar to the case involving the parabolic spectrum, 
the Green's function in the long-time limit has a power-law decay in time implying the breakdown of the quasiparticle description of the impurity. The spectral function in turn exhibits a weak power-law singularity.
In the recoil case, however, the reduced phase-space for scattering processes implies a non-zero quasiparticle weight and the presence of a coherent part in the spectral function.  
Performing a weak coupling analysis we find that the 
mobility of the impurity reveals a $T^{-3/2} $ divergence at low temperatures.
In addition, we show that the  Green's function of an impurity interacting with the helical edge modes (surface states of 2D TI),  
exhibit power-law decay in the long-time limit for both the non-recoil and recoil case (with low impurity momentum), indicating the break down of the quasiparticle picture.  However, for impurity with high momentum, the quasiparticle picture is restored. Using the Boltzmann approach we show that the presence of the magnetic field results in a power-law divergence of the impurity mobility at low-temperatures.

\end{abstract}

\maketitle


\section{Introduction} {\label{sec:I}}
In a pioneering work, Anderson (1967) showed that in  the thermodynamic limit adding an impurity as a perturbation in a many-particle system consisting of free electron gas results in a vanishing overlap between the initial unperturbed ground state and the final ground-state leading to a phenomenon known as the orthogonality catastrophe (OC)\cite{anderson1967}.
This idea was extended by  Nozieres and De Dominicis  into a dynamical theory of the  absorption process by studying the long-time behavior of the core-hole Green's function\cite{noz1969}.
A  number of studies have uncovered drastic modifications to the fermionic system due to its interaction with a single impurity. Some of the examples include, the x-ray edge effect~\cite{Mahan1967,schotte1969,gunnar1978,ohtaka1983,leggett1987}, the Kondo problem~\cite{yuval1970,Kondo1983,Kagan1992,latta2011}, impurity in a  semiconductor quantum dot~\cite{hakan2011,wolfgang2012},  impurity interacting with a Luttinger
liquid~\cite{gogolin1993,pusti2006}, heavy particle in a fermionic bath~\cite{OhtakaRMP,Prokofev1993,rosch1995,rosch1998} and impurity interaction with the fermionic many-body environment in ultracold atomic systems~\cite{demler2011,fukuhara2011,knap2012,fukuhara2013,Schmidt_2018}.

Lately, a new class of materials, the  3D and 2D topological insulators (TI) having  unusual surface states has  generated tremendous interest~\cite{HK2010,ZQ2011}.The TIs exhibit insulating behavior in the bulk but have surface states which are metallic 
and  are described by the relativistic Dirac equation~\cite{kane2005,bernevig2006,bernevigN2006,roy2006,fu2007,fuprb2007, moore2007,qi2008}.
They have an odd number of gapless Dirac-cones in which the spin and momentum are locked together into a  spin-helical state and are protected by the time-reversal symmetry (TRS).   The physics of  3D TIs has been studied in quite detail, some of  the questions addressed include magnetoelectric response in TI~\cite{Essin2009,Qi2009,Maciejko2010}, integer quantum hall effect~\cite{DHLee2009},  competition between localization and anti-localization~\cite{Hai-Zhou2011,Garate2012}, the  effects of  phonon and disorder on transport~\cite{Qiuzi2012},  bulk-surface coupling~\cite{Kush2014}, impurity dynamics at the particle-hole symmetric point~\cite{Caracanhas}, 
role of magnetic and nonmagnetic impurities from the point of view of their effect on local charge/spin density of states and also on the surface states of 3D TI  with Dirac spectrum has been an extremely active area of research \cite{liu2009,chen2010,biswas2010,he2011,zhu2011,annicaR2012,annica2012,qaium2012,lee2013,annica2014,ochoa2015,zyuzin2016,bobkova2016} etc. At the same time, a number of work on the interacting surface states of 2D TI which are  the helical Luttinger liquid  have been made. These include studies on the Kondo effect in the helical edge liquid~\cite{Maciejko2009},  Coulomb drag~\cite{Zyuzin2010}, spin susceptibility~\cite{Klinovaja2013,Meng2013}, transport~\cite{Schmidt2013,Meng2014},  structure factor~\cite{suh2014},  the role of  inelastic scattering channels on transport~\cite{Schmidt2012,Jan2012,Maestro2013},
etc.

In this work, we consider the interaction of Dirac fermions  with a nonmagnetic mobile impurity.
The unusual surface states of TI  provide an intriguing new scenario for the study of the phenomenon of orthogonality catastrophe in these systems.
We find that similar to that in a fermionic bath~\cite{Prokofev1993,rosch1995,rosch1998},  the physics of the heavy particle in a bath of Dirac fermions is strongly influenced by the presence or absence of infrared singularity. In the $D=2$ the interaction between the  bath and  a particle with the recoilless   mass generates an infinite number of low-energy particle-hole pairs resulting in an incoherent behavior  of the heavy particle, i.e., the quasiparticle weight vanishes. The spectral 
function, in turn, exhibits a  power-law divergence at the renormalized energy.
In contrast,  the recoil of the heavy particle suppresses the phase space available for particle-hole generation resulting in non-zero quasi-particle weight and consequently a $\delta$-function peak in the spectral-function.
However, a part of the spectral-weight is transferred to the incoherent part which exhibits a square-root singularity. We find that the Maxwell-Boltzmann distribution of the mobile impurity governs the typical momentum transfer between the impurity and the  Dirac fermions resulting in a $T^{-3/2}$ temperature dependence of the mobility of the impurity.

The study of interaction effects between the mobile
impurity and 1D helical Luttinger liquid reveals that
the quasiparticle weight vanishes except for the scenario
when the momentum of the mobile impurity with mass
$M$ exceeds $M v$, where $v$ is the sound velocity. This result is in agreement with an earlier study of a single spin-down fermion interacting with the bath of spin-up fermions~\cite{Kantian2014}.   As for mobility, in the absence of a magnetic field, the mobility of the impurity is limited by forwarding scattering-processes only and diverges exponentially.
However, turning on the magnetic-field results in a power-law divergence at low-temperatures.

The  paper is organized as follows: Section~\ref{sec:II}  includes
a general description of our model along with the Green's function of the Dirac fermions in TI for the 2D case.
In section~\ref{sec:III} the linked cluster technique has been used to develop a formalism for obtaining  the Green's function of  impurity interacting with the Dirac fermions. In addition, the long-time behavior of the impurity Green's function for the recoilless and the recoil case and the  corresponding spectral-function have been  studied.  In section~\ref{sec:IV} the temperature dependence of the  mobility of  the impurity has been obtained.  In  section~\ref{sec:V} we establish the model for the 1D case  and discuss impurity Green's function for the recoilless and the recoil case. The mobility of impurity interacting with 1D helical liquid is discussed in detail in section~\ref{sec:VI} followed by a section on the summary of our results. 

 
 \begin{figure}
\centering
\includegraphics[width=1.0\linewidth]{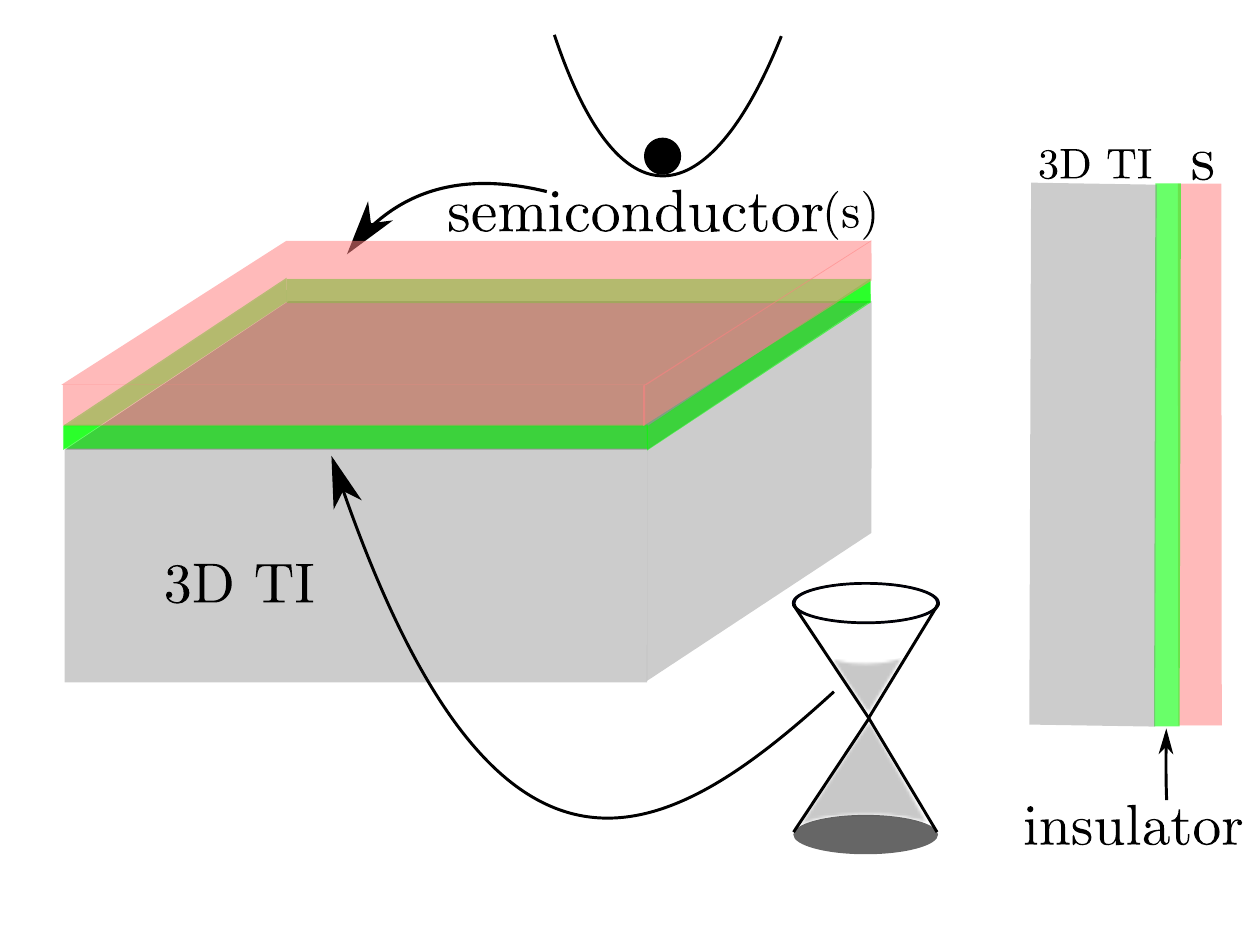}
\caption{(Color online) Schematic picture of our model where we  consider a semiconductor placed on  top of the surface of a  3D TI  with a  conventional insulator in between them. We assume a very large energy barrier for the electrons to hop on either side.  A single mobile electron in the conduction band of the  semiconductor (with completely filled valence band)   behaves like an impurity in the environment of Dirac fermions.}
\label{model_fig}
 \end{figure}

 \section{Model for the 2D Case} {\label{sec:II}}
 We  consider the motion of a heavy particle with mass $M$ having a parabolic dispersion (for example in a 2D-semiconductor) and constrained to move in two-dimensions. The semiconductor is placed on top of the surface of a 3D topological insulator separated by a thin insulating layer (see Fig.~\ref{model_fig}). We make following three assumptions:  the bulk is insulating and does not influence the physics, absence of tunneling  between the TI and the semiconductor  and that the  heavy mass $M$ interacts with the    Dirac fermions 
 via a contact potential. The low energy effective Hamiltonian of the 2D Dirac fermions has the following form~\cite{liu2014,Zhou2017} $H_{D} =\hbar v_F(\hat{z}\times\hat{k})\cdot \hat{\tau}$, where $\tau$'s are the Pauli matrices. Performing a  simple unitary transformation the Hamiltonian of 
 this composite system can be written as
 \begin{eqnarray}
 H=\frac{p^2}{2M}+\hbar v_F(\vec{\sigma}\cdot\vec{k}) +\sum\limits_{q} V(q)\rho(q)n(-q),
 \label{ini_H1}
 \end{eqnarray} 
 where $p$ is the momentum of the particle and  the second term  represents the transformed low-energy effective Hamiltonian of the Dirac fermions. 
 Henceforth we will work in the $\hbar=1$ and $v_F=1$ units unless  specified otherwise. The third term is the interaction term, 
 where the potential $V(q)=U/A$ is momentum independent and  the density operators $\rho(q)$ and $n(q)$ correspond to the  Dirac fermions and the impurity particle, respectively.
 The second quantized form of the Hamiltonian in Eq.(\ref{ini_H1}) acquires the following form
 \begin{eqnarray}
 &&H=\sum_{P}\epsilon_p\hat{a}^{\dagger}_p\hat{a}_p + \sum_{k,\alpha,\beta}\hat{c}_{k\alpha}^\dagger (\vec{\sigma}\cdot\vec{k})\hat{c}_{k\beta} + V, \notag
 \label{ini_H2}
 \end{eqnarray}
 where $\epsilon_p=p^2/2M$ is the energy of the particle and the interaction potential in the second quantized notation is 
 \begin{eqnarray}
 V= U\sum_{\sigma, k_1,k_2,q}  \hat{a}_{k_2-q}^{\dagger}\hat{a}_{k_2} \hat{c}_{k_1+q,\sigma}^{\dagger}\hat{c}_{k_1,\sigma} .
 \end{eqnarray}
 The corresponding zero temperature Matsubara Green's function for the Dirac fermions on the surface of a 3D TI has the following form,
 \begin{equation}
 \mathcal{G}(k,i\omega)=\frac{1}{2}\sum_{\eta=\pm 1}\left[\frac{\hat{I}+\eta(\vec{\sigma}\cdot \vec{\bar{k}})/\xi_k}{i\omega-\eta\,\xi_k+\mu_F}\right],
 \end{equation}	
 where $\vec{\bar{k}}=k_x\hat{e}_1+k_y\hat{e}_2+\Delta \hat{e}_3$ and $\xi_k=\sqrt{k^2 +\Delta^2}$ is the dispersion relation of Dirac fermions and $\Delta$ is the mass term which opens up a gap in the TI. Considering the Dirac fermions in the upper band only, 
 the expression for the Green's function in the momentum-time representation is given by
  \begin{eqnarray}
 \hat{\mathcal{G}}(k,t)=\frac{\Big[\hat{I}+\frac{ (\vec{\sigma}\cdot\vec{\bar{k}})_{\alpha \beta}}{\xi_k}\Big]}{2i}\bigg[\theta(t)(1-n_k)-\theta(-t)n_k\bigg]e^{-i\bar{\xi}_k t},\qquad
 \end{eqnarray}
 where $\bar{\xi}_k=\xi_k-\mu_F$.
 
\section{Impurity Green Function}{\label{sec:III}}

In the following, we will utilize the linked cluster  method to obtain the expression for  the Green's function of an impurity particle interacting with the surface states of a 3D TI.
The approach is similar to the one used for the polaron problem, here instead, we will incorporate the interaction between the impurity particle and the Dirac fermions. The expression for the impurity Green's function to all orders in interaction has the following form:

\begin{figure}
	\centering
	\includegraphics[width=1.0\linewidth]{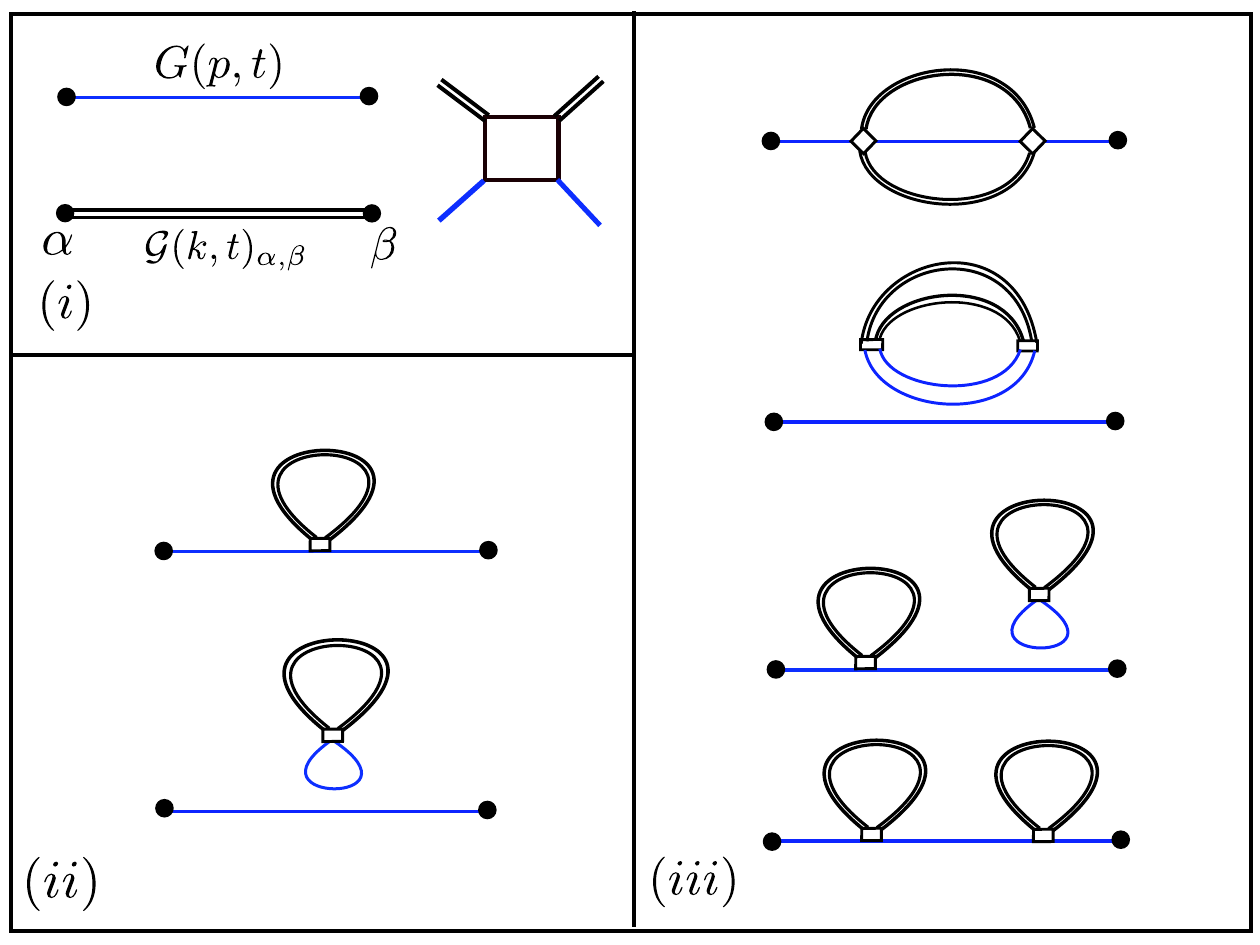}
	\caption{(Color online)  (i) The bare propagator and the  interaction vertex.  (ii) First order, and (iii) second order diagrams. Only the connected diagrams are relevant for the impurity Green's function calculation.}
	\label{Diag}
\end{figure}
\begin{eqnarray}
	 G(p,t) =\sum_{n=0}^{\infty} \mathcal{M}_n (p,t),
\end{eqnarray}
where
\begin{eqnarray}
\mathcal{M}_{n}(p,t)= \frac{(-i)^{n+1}}{n!} \int_{0}^{t}dt_1 \cdot\cdot\cdot\int_{0}^{t}dt_n\mathcal{C}_n,\quad
\end{eqnarray}
and  
\begin{eqnarray}
\mathcal{C}_n= \langle\hat{a}_p (t) V(t_1)\cdot\cdot\cdot V(t_n)\hat{a}_{p}^\dagger (0)\rangle.
\end{eqnarray}
In terms of the cumulants, $\mathcal{S}_n$, the Green's function can be  re-expressed as 
\begin{eqnarray}
G(p,t)=G_0\exp\big[\sum_{n=1}^{\infty} \mathcal{S}_n (p,t)\big],
\end{eqnarray} 
where 
$G_0 (p,t)= -i\Theta(t)\exp(-i\epsilon_p t)$ is the free Green's function of the impurity  particle. The dominant  contributions to the Green's function is already contained in the first two  cumulants given by\cite{Mahan1967,hart1971,Prokofev1993,rosch1995,rosch1998}
\begin{eqnarray}
\mathcal{S}_1=G_0^{-1}(p,t)\mathcal{M}_1,~\text{and}~ \mathcal{S}_2=G_0^{-1}(p,t)\mathcal{M}_2-\frac{1}{2!} \mathcal{S}_1 ^2.\notag
\end{eqnarray}
We first evaluate the  $\mathcal{M}_1$ term, 
\begin{eqnarray}
&&\mathcal{M}_1=(-i)^2\int_{0}^{t}dt_1 \mathcal{C}_1,\notag
\end{eqnarray}
where
\begin{eqnarray}
&&\mathcal{C}_1    
=U \sum_{k_1,k_2,q} \big\langle\mathcal{T}     
\hat{a}_p(t)\hat{a}_{k_1+q}^\dagger(t_1)\big\rangle\big\langle\mathcal{T}     
\hat{a}_{k_1}(t_1)\hat{a}_{p}^\dagger(0)\big\rangle\notag\\
&&\times~\big\langle\mathcal{T}      \hat{c}_{k_2-q,\sigma}^{\dagger}(t1)\hat{c}_{k_2,\sigma} (t1)\big\rangle.\label{C1}
\end{eqnarray}
In the above expression for $\mathcal{C}_1$, we employ the  Wick's theorem to decompose the averages involving more than two fermionic operators into products of bare Green's function. 
 
Note that  only the connected diagram  has been included (see Fig.~\ref{Diag}). The first two terms of Eqn.~\ref{C1} represent the  bare impurity Green's function  and the last commutator  gives us  the occupation number for Dirac fermions. Performing the integration over time we obtain, 
$\mathcal{M}_1=-iU\sum_{k_1}G_0 (p,t)n_{k_1}t$, where $n_{k_1}$ is the Fermi distribution function.  
Thus the first  cumulant is  
\begin{eqnarray}
\mathcal{S}_1=\frac{\mathcal{M}_1}{G_0}=-i\frac{U}{A}\sum_{k_1}n_{k_1}t.
\end{eqnarray}
The second cumulant is obtained from the  $\mathcal{M}_2$ term which involves scattering at two different times
\begin{eqnarray}
\mathcal{M}_2=\frac{(-i)^3}{2!}\int_{0}^{t}dt_1\int_{0}^{t}dt_2~ \mathcal{C}_2,\notag
\end{eqnarray}
where,	
\begin{widetext}
\begin{eqnarray}
\mathcal{C}_2 =\frac{U^2}{A^2}\sum
\Big\langle\mathcal{T}\Big\{a_p(t)a^{\dagger}_{k_2 -q}(t_1)a_{k_2}(t_1)c^{\dagger}_{k_1 +q,\sigma_1}(t_1) c_{k_1,\sigma_1}(t_1)a^{\dagger}_{k_4  -\bar{q}}(t_2)a_{k_4}(t_2)c^{\dagger}_{k_3 +\bar{q},\sigma_2}(t_2)
c_{k_3,\sigma_2}(t_2)a_p^\dagger(0)\Big\}\Big\rangle.
\end{eqnarray}
As before, we will use the  Wick's theorem to  simplify the above expression.
At the outset we will disregard the disconnected diagrams and also the terms which are obtained from squaring the first cumulant  (see Fig.~\ref{Diag}). The Dirac fermion commutators yield
\begin{eqnarray}
\big\langle\mathcal{T}\left\{c^{\dagger}_{k_1 +q,\sigma_1}(t_1)c_{k_1,\sigma_1}(t_1)c^{\dagger}_{k_3 +\bar{q},\sigma_2}(t_2)c_{k_3,\sigma_2}(t_2)\right\}\big\rangle\notag\hspace{0.25cm}
=\delta_{k_1,k_3+\bar{q}}\delta_{k_3,k_1+q}\text{Tr}\Big[\hat{\mathcal{G}}(k_1,t_1-t_2)\hat{\mathcal{G}}(k_3,t_2-t_1)\Big],\notag
\end{eqnarray}	
and from the impurity creation and annihilation operators we obtain:
\begin{eqnarray}
\mathcal{Z}=	\Big\langle\mathcal{T}\Big\{a_p(t)a^{\dagger}_{k_2 -q}(t_1)a_{k_2}(t_1)a^{\dagger}_{k_4  -\bar{q}}(t_2)a_{k_4}(t_2)a_p^\dagger(0)\Big\}\Big\rangle=e^{-i\epsilon_pt}\sum_{\eta=\pm}\Theta[\eta(t_1-t_2)]\exp\Big[ i\eta(\epsilon_p-\epsilon_{p+\eta q})(t_1 -t_2)    \Big]
	\notag.
	\end{eqnarray}
Thus  $\mathcal{S}_2$ is given by
\begin{eqnarray}
\mathcal{S}_2=\frac{G_0^{-1}(p,t)}{2}\frac{U^2}{A^2}\sum_{k_1,k_2}\int dt_1 dt_2\,\mathcal{Z}\,\,\text{Tr}\Big[\mathcal{G}_{k_1}(t_1-t_2)\mathcal{G}_{k_2}(t_2-t_1)\Big]\notag.
\end{eqnarray}
Performing  the integration on time, we obtain
\begin{eqnarray}
\mathcal{S}_2=\frac{U^2}{A^2}\sum_{k_1 k_2}\Big[1+\hat{k}_1\cdot\hat{k}_2\Big](1-n_{k_2})n_{k_1}\Bigg[\frac{it}{\tilde{\Delta}}
-\frac{1-e^{-i\tilde{\Delta}t}}{\tilde{\Delta}^2}\Bigg],\label{cumu2}
\end{eqnarray}
where,  $\tilde{\Delta}(k_1,k_2) = \epsilon_{p+k_1-k_2}-\epsilon_p +\xi_{k_2}-\xi_{k_1}$.  
Note that the chiral form in~(\ref{cumu2}) is a feature of the particle-hole pairs in the Dirac sea.  
Putting together $\mathcal{S}_1$ and $\mathcal{S}_2$ we obtain the following expression for the impurity Green's function 
\begin{eqnarray}
iG(p,t)=\Theta(t) \exp \big[ -i\tilde{\epsilon}_p t+ \mathcal{X}(t)\big],\hspace{0.95cm}
\label{green_I}
\end{eqnarray}
where the renormalized energy $\tilde{\epsilon}_p$ is given by
$$\tilde{\epsilon}_p=\epsilon_p+\frac{U}{A}\sum_{k}n_k  - \frac{U^2}{A^2}\sum_{k_1,k_2}\Big[1+\hat{k}_1\cdot\hat{k}_2\Big]\frac{(1-n_{k_2})n_{k_1}}{\tilde{\Delta}(k_1,k_2)},$$
while the function $\mathcal{X}(t)$ which will be our object of interest encodes the non-trivial $t$ dependence and is given by
\begin{eqnarray}
\mathcal{X}(t) = -\frac{U^2}{A^2}\sum_{k_1 k_2}\Big[1+\hat{k}_1\cdot\hat{k}_2\Big](1-n_{k_2})n_{k_1}\frac{1-e^{-i\tilde{\Delta}t}}{\tilde{\Delta}^2}.\hspace{0.95cm}
\label{Xt}
\end{eqnarray}
The following  change of variables: $k_1 \rightarrow k$ and $k_2 \rightarrow k+q=k_q$, 
allows us to  rewrite  $\mathcal{X}(t)$ in the following  compact form
\begin{eqnarray}
\mathcal{X}(t) = \frac{U^2}{ A}\sum_q\int \frac{d\omega}{\pi}~\text{Im}\Pi(q,\omega-\epsilon_{p+k-k_q}+\epsilon_p ) \frac{1-e^{-i\omega t}}{\omega^2},
\label{Xt1}
\end{eqnarray}
where the imaginary part of the  zero-temperature polarization operator $\text{Im}\Pi(q,\omega) $ is given by
\begin{eqnarray}
{\text{Im}}\Pi(q,\omega)&=&-\frac{\pi}{A}\sum_{k} \left[1+\hat{k}\cdot \hat{
k}_q\right]n_{k}\left[1-n_{k_q}\right]\delta(\omega-\xi_{k+q}+\xi_{k}).\label{ImPi}
\end{eqnarray}
We will make use of the expressions given in Eqs.~(\ref{Xt1})
and (\ref{ImPi}) to evaluate the behavior of Green's function in the limiting case of infinite mass and for the finite mass scenario.
\end{widetext}

\subsection{Infinite mass and Non-recoil of Impurity}

In the  limit of heavy mass, $\tilde{\Delta}$ can be approximated as  $ \xi_{k_q}-\xi_{k}$ and  the polarization term in~(\ref{Xt1}) as $~\text{Im}\Pi(q,\omega) $.  
Since our primary goal is to obtain the behavior of the Green's function in the long-time limit, 
it suffices to consider the momentum integration $\rho(\omega)= \int d^2q/(2\pi)^2\text{Im}\Pi(q,\omega)$, arising from the low-frequency regime   of  the polarization function.  
 
\begin{figure}
\centering
\includegraphics[width=1.0\linewidth]{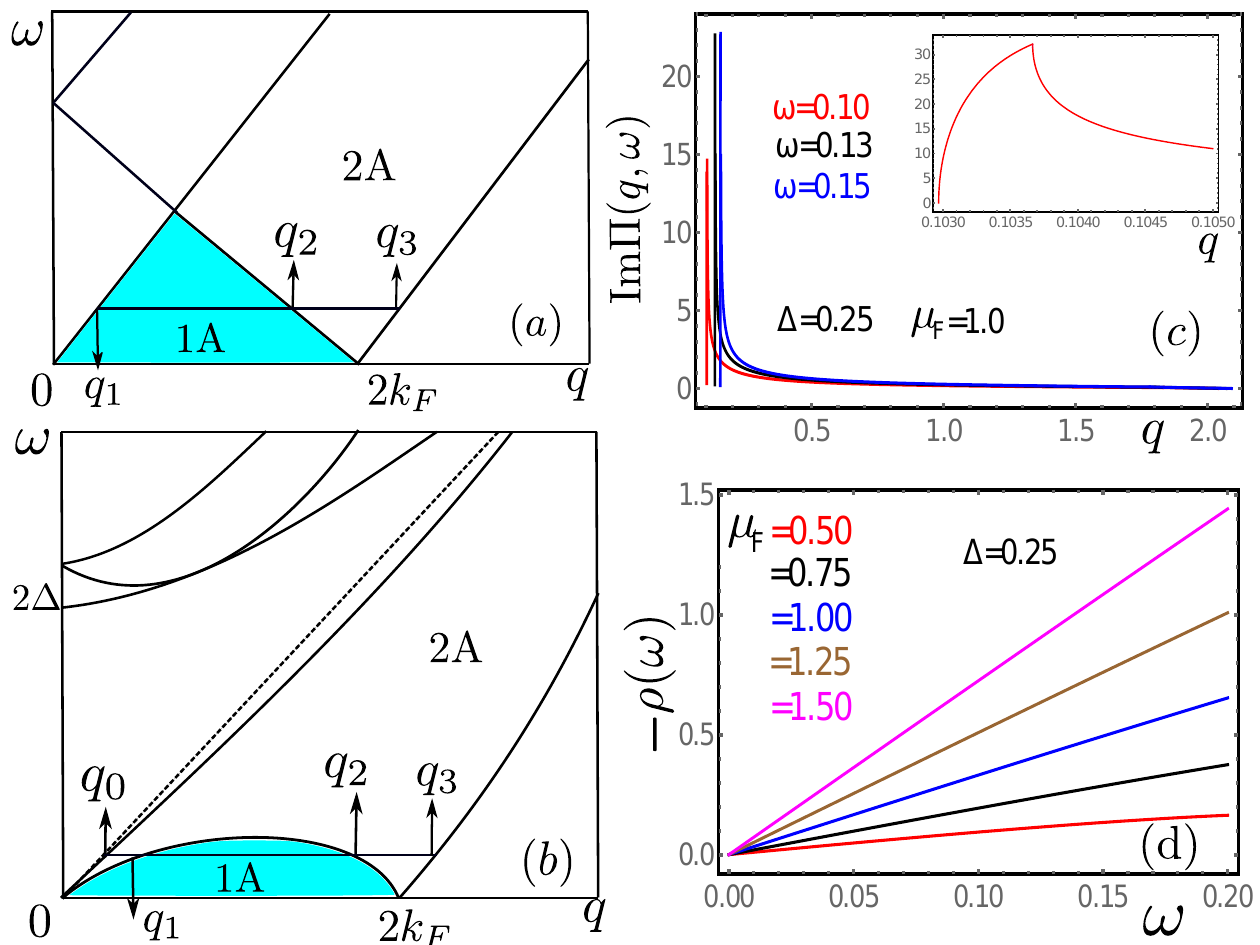}
\caption{(Color online)  Non-recoil scenario: the allowed particle-hole regions in the $(q,\omega)$ plane.~(a)~For gapless excitation.~(b)~For $\Delta\ne 0$ and $\Delta\ll\mu_F$.  Here  $q_{0/1}=\mp k_F\pm\sqrt{(\mu_F\pm\omega)^2-\Delta^2}$ and $q_{2/3}=k_F+\sqrt{(\mu_F\mp\omega)^2-\Delta^2}$.  (c) The imaginary part of charge susceptibility  $\text{Im}\Pi(q,\omega)$ as a function of $q$. The inset shows  $\text{Im}\Pi(q,\omega)$ for $\omega = 0.1$ and small values of $q$. (d) Linear behavior of $\rho(\omega)$ as a function of $\omega$ for a range of $\mu_F$ and fixed $\Delta$.}
\label{allowed_region}
\end{figure}

We will split the  integration in to three regions and explicitly compare their contributions. In the regions $q_0<q<q_1$ and $q_2<q<q_3$  shown in  Fig.~(\ref{allowed_region}),  the  polarization operator 
has the following  form~\cite{Sarkar}
\begin{eqnarray}\label{ImPiB}
\text{Im}\Pi (q,\omega)=-\frac{1}{2\pi\sqrt{q^2-\omega^2}}\bigg[\mathcal{F}(2\mu_F+\omega)-\mathcal{F}(\zeta)\bigg], \qquad
\end{eqnarray}
where $\mathcal{F}(x)$ and $\zeta$ are as given in the appendix~(\ref{pol_func}).

The evaluation of the integral in the first region yields $-2  \omega^{5/2}\Delta/\pi^2\sqrt{k_F}$ while the $q_2<q<q_3$ 
region yields a larger contribution $ -(16\omega^{3/2}\sqrt{k_F}/3\pi^2) \text{max}[\Delta^2 /\sqrt{2}k_F,\omega/5]$.

The largest contribution is obtained from the $q_1<q<q_2$ region wherein the polarization function is given by
\begin{eqnarray}\label{ImPiA}
\text{Im}\Pi_A (q,\omega)=- \frac{1}{2\pi\sqrt{q^2-\omega^2}}\sum_{\eta=\pm}\eta\mathcal{F}(2\mu_F+ \eta\omega).\qquad
\end{eqnarray}
The leading term obtained upon the momentum integration yields  linear in $\omega$ term given by 

\begin{eqnarray}
&&\int \frac{qdq}{2\pi}\text{Im}\Pi(q,\omega) \approx  -\frac{\omega}{2\pi^2}\int_{q_1}^{q_2}  qdq\frac{\partial \mathcal{F}(x)}{\partial x }\Big{|}_{x=2\mu_F}\notag\\
&&=-\frac{1}{\pi^2} \int_{q_1}^{q_2} qdq\frac{\omega (4\mu_F^2-q^2)}{\sqrt{4\mu_F^2 (q^2-\omega^2) -q^2 (4 \Delta^2 + q^2 -\omega^2) }}\notag\\
&&\approx-\frac{k_{F}^2\omega}{\pi}.
\end{eqnarray}
In addition, we obtain a second     linear in $\omega$ term, which however,  is smaller by a factor of $\Delta^2/\mu_F^2$. Keeping the dominant term in $\mathcal{\chi}(t)$ and in the long-time limit we obtain~\cite{hart1971}
\begin{eqnarray}
\mathcal{\chi}(t)
\approx -\frac{k_{F}^2 U^2}{\pi^2}\log(1+it\omega_c),
\end{eqnarray}
where $\omega_c$ is the bandwidth and is taken to be of the order of  Fermi-energy.
Thus the  behavior of the Green's function~(\ref{green_I}) in the long time limit is determined by the  $t$ and the  $\log t$ term both of which are in the exponential.
The latter term leads to  a power-law decay of the Green's function  $\propto 1/t^{\nu}$, where $\nu=k_F^2 U^2/\pi^2$ and   is responsible for the orthogonality catastrophe.\\  

Besides the  Green's function, the spectral function of the heavy particle acquires drastic modification as compared to the free case. The spectral function is given by
\begin{eqnarray}
\mathcal{A}(\epsilon)=-2\,{\text{Im}}\bigg[\int_{-\infty}^{\infty}
dt e^{i\epsilon t}G(t)\bigg]=\frac{e^{-\tilde{\epsilon}}}{i\mu_F}\int_{1-i \infty}^{1+i\infty}dz \frac{e^{{z\tilde{\omega} }}}{z^{\nu}}\notag,
\end{eqnarray}
where $\tilde{\epsilon}=(\epsilon-\tilde{\epsilon}_p)/\mu_F$.  First consider the case $\tilde{\epsilon}<0$, since $e^{{z\tilde{\epsilon} }}/z^{\nu}$ is analytic everywhere for $Re(z) >1$,
the  contour of integration can be pushed to $Re(z) >1$ and $|z|\rightarrow\infty$. The integrand vanishes everywhere for the modified contour, therefore $\mathcal{A}(\epsilon)=0$  for $\tilde{\epsilon}<0$. 
On the other hand, for $\tilde{\epsilon}>0$, the integrand is analytic everywhere  except for the negative real axis where it has a branch cut.
Therefore, the contour can be deformed on to the  negative real axis and we obtain
\begin{eqnarray}
\mathcal{A}(\epsilon)=\frac{2}{\mu_F}\text{Im}\big[\int_0^\infty dr r^{-\nu} e^{-r\tilde{\epsilon}} e^{i\pi\nu}\big]=\Theta(\tilde{\epsilon})\frac{2\pi}{\mu_F}\frac{e^{-\tilde{\epsilon}}\tilde{\epsilon}^{\nu-1}}{\Gamma(\nu)}.\notag\\
\end{eqnarray}
Thus the spectral function is no longer a delta-function  peaked at the renormalized energy $\tilde{\epsilon}_p$, instead
due to the large number of particle-hole excitations has a 
power-law singularity given by $\mathcal{A}(\epsilon) \propto \Theta(\epsilon-\tilde{\epsilon}_p)/(\epsilon-\tilde{\epsilon}_p)^{1-\nu}$. Thus the localized impurity acts as an incoherent excitation due to its interaction with the Dirac electrons and  decays with time. \\ 


\subsection{Recoil Case: Suppression of Orthogonality Catastrophe}
The above-discussed scenario is significantly modified when considering an impurity with finite mass. In a typical scattering event involving an  impurity atom and  a particle-hole pair 
with momentum $q$ and energy $\omega$ (where $q v_F\gtrsim \omega$) the impurity momentum changes by $q\sim\sqrt{2M \omega}$. 
Thus   for $\sqrt{2M \omega} \ll 2k_F$, the  phase-space available for low-energy scattering  is severely restricted.
This, in turn, is reflected in the deviation of $\rho(\omega)$ from the linear behavior and results in a modified Green's function. 
  
\begin{figure}
\centering
\includegraphics[width=1.0\linewidth]{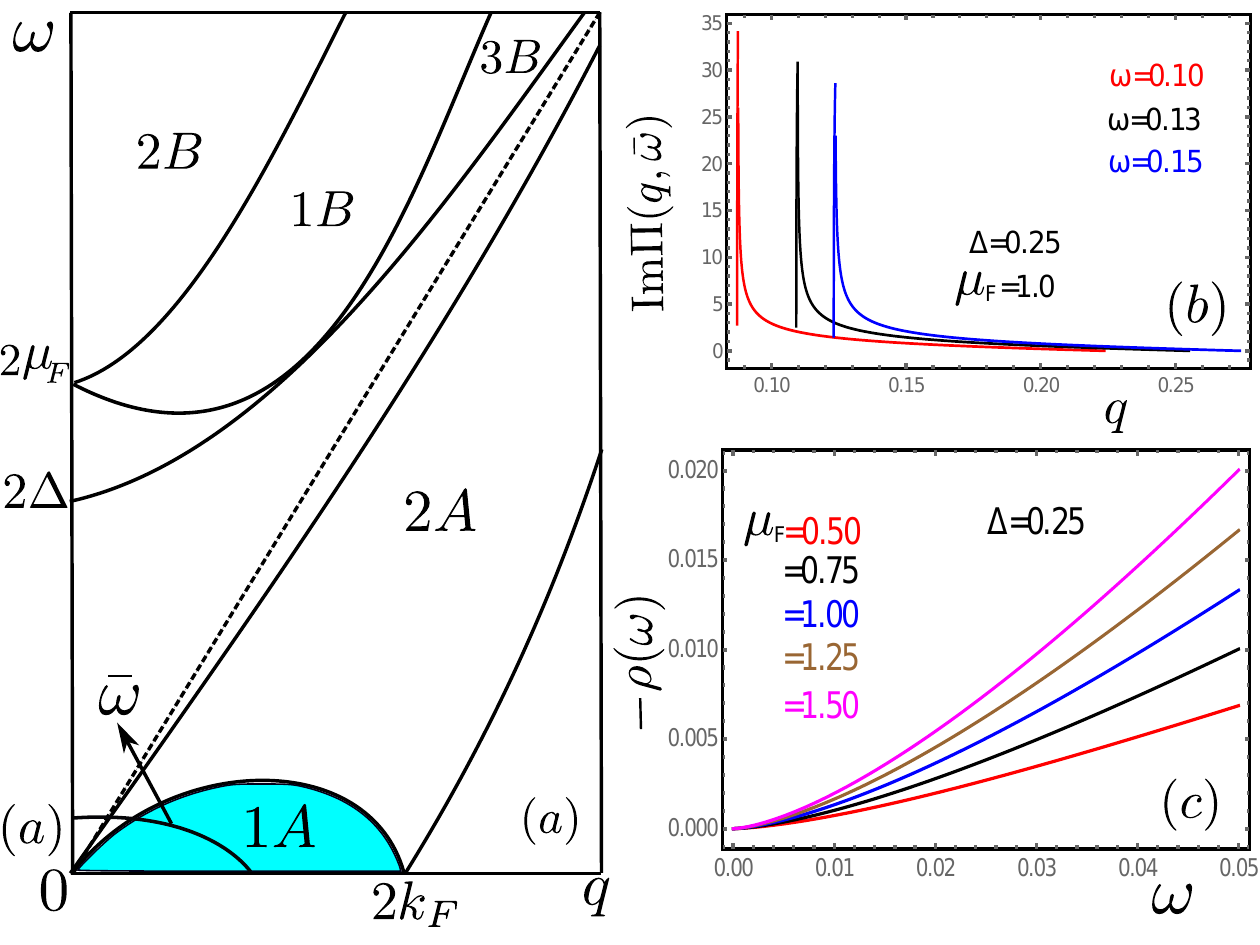}
\caption{(Color online)  
	Recoil scenario: (a) The particle-hole excitation regions 
	for gapped excitation of Dirac fermions and  $\tilde{\omega} =\omega -q^2/2M$. (b) Plot of $\text{Im}\Pi(q,\tilde{\omega})$ as a function of $q$ for fixed values of $\omega$. (c) The $\omega^{3/2}$ behavior of $\rho(\omega)$ for different $\mu_F$ and fixed $\Delta$. }
\label{allowed_region2}
 \end{figure}

Following earlier discussion, $\rho(\omega)$ for the recoil case is given by,
\begin{eqnarray}\label{recoil_DOS1}
\rho(\omega)=\int\frac{d^2 q}{4\pi^2}\,\,\text{Im}\Pi(q,\omega- \epsilon_{\vec{p} + \vec{q}}  +\epsilon_{ \vec{p}}  ).
\end{eqnarray}
In the limit of small frequency and vanishingly small momentum of the impurity, the limits of integration (see Fig.~\ref{allowed_region2}) are from  $\omega$ to $\sqrt{2M\omega}$, where $\sqrt{2M\omega} \ll 2k_F  $ and we have assumed $\Delta\ll \mu_F$. Using the expression for  the polarization operator given in Eq.~(\ref{App_ImPi}) and replacing $\omega\rightarrow \omega -q^2/2m$,  $\rho(\omega)$ acquires the following form,
\begin{eqnarray}\label{recoil_DOS2}
\rho(\omega)=-\frac{1 }{\pi^2}\int_{\omega}^{\sqrt{2M\omega}} dq \frac{(\omega-\frac{q^2}{2M}) q}{\sqrt{q^2-(\omega-\frac{q^2}{2M})^2}}\Bigg[\frac{4\mu_F^2-q^2}{\sqrt{4\mu_F^2-\zeta^2}}\Bigg],\notag
\end{eqnarray}	
where the leading order result is given by 
$\rho(\omega)= -g  \omega^{3/2}$, with the proportionality constant being $g= \frac{4\sqrt{2M}}{3\pi^2}k_F$. Thus compared to the infinite mass scenario, recoil of the impurity causes
suppression of the particle-hole excitation and as will be shown below  the impurity quasiparticle weight remains non-zero.

The quasiparticle weight $\text{Z}_0$ is obtained from evaluating the time independent part  of $\mathcal{X}(t)$~(\ref{Xt1}), i.e.,
$$U^2\int \frac{d\omega}{\pi}~ \frac{\rho(\omega)}{\omega^2},$$
yielding $\text{Z}_0\approx \exp[-2gU^2\sqrt{\omega_c}/\pi] $. 
As in the infinite mass case the linear in time  term in the exponential,  $\exp(-i\epsilon_p t)$,
gets trivially renormalized to
$\bar{\epsilon}_p$.
However,  unlike the log term which is responsible for the 
strong suppression of the Green's function of the infinite mass, here the time-dependent integral of $\mathcal{X}(t)$ results in a $t^{-1/2}$ term,  specifically  
$$-U^2\int \frac{d\omega}{\pi}~ \frac{\rho(\omega)}{\omega^2}e^{-i\omega t}\approx  gU^2\frac{e^{-i\pi/4}}{\sqrt{\pi}\sqrt{t}}\approx  \frac{gU^2e^{-i\pi/4}}{\sqrt{\pi}}t^{-1/2}.  $$

Therefore the long-time behavior of  the Green's function acquires the form
\begin{align}
G(p,t)= -i\Theta(t)\text{Z}_0\exp\left(-i\bar{\epsilon}_p t+\frac{gU^2e^{-i\pi/4}}{\sqrt{\pi}}t^{-1/2}\right).
\end{align}
We note that for $t\rightarrow\infty$ the second term in the exponential vanishes and 
we are left with a Green function describing a well defined quasiparticle excitation with $Z_0<1$.

As before, an insightful perspective into the nature of excitations is revealed from the behavior of the spectral function $\mathcal{A}(\epsilon)$.
The small contribution to the Green's function due to the $t^{-1/2}$ term allows for a perturbative treatment of the spectral function. Therefore, the spectral function can be split into a coherent and incoherent part.  The coherent part is given by $\mathcal{A}^{\text{Coh.}}(\epsilon) \approx \text{Z}_0\delta(\epsilon-\bar{\epsilon}_p)$.
On the other hand, the incoherent part  has a square-root singularity with the following expression, 

\begin{eqnarray}
\mathcal{A}^{\text{Incoh.}}(\epsilon) \approx 2  g U^2 \text{Z}_0 \frac{\Theta(\epsilon-\bar{\epsilon}_p)}{\sqrt{\epsilon-\bar{\epsilon}_p}},\label{incoh}
\end{eqnarray}
and  is obtained by  performing a partial series expansion of the above Green's function and taking the imaginary part of the Fourier transform of $\delta G \propto -i\Theta(t)e^{-i\bar{\epsilon}_pt}t^{-1/2}$, where   we have made  use of  the following result:  $\int_0^\infty e^{i\alpha t}dt/\sqrt{t}=\sqrt{\pi}e^{i\text{sgn}(\alpha)\pi/4}/\sqrt{|\alpha|}$.

The non-zero quasiparticle weight and the delta-function in the spectral function attest to the well-behaved quasiparticle like excitation. At the same time, the weaker square-root singularity in the incoherent part is indicative of the remnants of the orthogonality physics that is significantly subdued due to the relatively fewer number of particle-hole excitations generated in the recoil process.

\section{Mobility of Impurity}{\label{sec:IV}}

In this section we will obtain  the low temperature behavior  of  the DC mobility which is given by  $\mu =e \tau/M $, where $\tau$ is the transport time~\cite{rosch1998}.  We estimate  $\tau$  by first calculating the inverse quasiparticle lifetime for a mobile impurity with momentum $p$ using the Fermi's golden rule \cite{Pines1}
\begin{eqnarray}
\frac{1}{\tau_p}=- \int&& \frac{d\omega d^2q}{2\pi^2}U^2(q)
\frac{1}{e^{\beta \omega}-1}
\text{Im}\Pi(q,\omega)\notag \\
&&\times~
 \delta(\omega+\epsilon_{p}-\epsilon_{p+q}) ,
\end{eqnarray}
where  the above expression is a modified version of the standard formula for the life-time of fermions which has an additional  $[1-n_F(\epsilon_{p+q})]$ factor. 
The  term represents the probability that  the scattered state is unoccupied,  which in our case is simply  set to unity as the  corresponding   impurity state remains unoccupied. The identical expression is obtained from the on-shell imaginary part of the self-energy of the mobile impurity. 

The statistical average of $1/\tau_p$ is  performed  with respect to the Boltzmann weight factor.  We denote the average as  $\langle 1/\tau_p \rangle$ given by

\begin{eqnarray}
\Big\langle\frac{1}{\tau_p}\Big\rangle=\frac{\beta}{2\pi M }\int d^2p~\frac{1}{\tau_p} e^{-\beta\epsilon_p}.\hspace{0.5cm}
\label{avglt}
\end{eqnarray}
For our purpose the above expression  is useful as the time-scale obtained from it  yields the same order of magnitude and the temperature dependence as the transport time. 

The energy scale in the integral of Eq.~(\ref{avglt}) is set by the temperature. Therefore, the  contribution to the integrals are dominated by the regions $p,q\sim \sqrt{2MT}$ and $\omega\sim T$. 
In the low temperature regime ($T\ll k_F^2/M$) the typical momentum transferred $q$ satisfies $q\ll k_F$, moreover, $\omega/qv_F\ll 1$ which implies   the polarization operator  can be expanded in the ratio $\omega/qv_F$  yielding 
$\text{Im}\Pi(q,\omega)\approx -(4/\pi)\mu_F\omega/qv_F$.
Performing the angular integration removes the  $\delta$-function and  yields

\begin{eqnarray}
\Big\langle\frac{1}{\tau_p}\Big\rangle=&&\frac{4\mu_F U^2_0}{\pi^3 v_F }(MT^3)^{1/2}\int_{0}^{\infty} \tilde{p}~d\tilde{p}    e^{-\tilde{p}^2/2} \int_{0}^{\infty} d\tilde{q}\notag \\&& \int_{-\tilde{p}^2/2}^{\infty}  d\tilde{\omega}
\frac{1}{e^{\tilde{\omega}}-1}
~\frac{\tilde{\omega}}{\sqrt{(\tilde{p}\tilde{q})^2-(\tilde{\omega} -\tilde{q}^2/2)^2}},\hspace{0.5cm}
\end{eqnarray}
where  we have used dimensionless variables
$\tilde{p}=p/\sqrt{MT}$, $\tilde{q}=q/\sqrt{MT}$ and $\tilde{\omega}=\omega/T$. The lower cut-off on the frequency integration is imposed by the $\delta$-function which forbids the frequency range  $\omega < -\epsilon_p$. We note that the dimensionless integral is of order $\mathcal{O}(1)$, while the  change of variables allows us to  extract the $T^{3/2}$ temperature dependence of the inverse scattering time. The above result emphasizes the fact that mobility of impurity interacting with Dirac fermions on the surface of TI in the low temperature region diverges  with decreasing temperature as $\mu \propto T^{-3/2}$.\\

\section{Interaction of impurity with the helical edge state}{\label{sec:V}}

So far we have considered the interaction of an isolated impurity with that of the surface states of a 3D TI. Similar to a 3D TI, a 2D TI 
has an insulating bulk and metallic edge states. 
The pair of gapless-edge states have specific chirality (also called helical edge-states) and are time-reversed partners of each other. These are the 1D helical modes in which backscattering due to the nonmagnetic impurities is forbidden.
A gap in the spectrum can be introduced by breaking time-reversal symmetry which is typically achieved by an external magnetic field.
In this section, we will first develop the formalism to describe the interaction of an isolated mobile impurity with that of an interacting helical liquid followed by the study of Green's function in the non-recoil and recoil case.

The non-interacting Hamiltonian of a helical liquid  in the presence of a magnetic field has the following form
\begin{align}\label{HL}
H_{\text{HL}}^0=\int dx\psi^\dagger(x)(-i\hbar \partial_x\sigma_z +B\sigma_x -\epsilon_F)\psi(x),
\end{align}
where $B$ is the Zeeman field applied along  the $x$-direction  (taken to be perpendicular to the spin-quantization axis) and 
the dispersion 
is given by $\epsilon_{\pm}=\pm\sqrt{v^2p_x^2+B^2}$.  $\gamma_p=\tan^{-1}(p/B)$, while  $\hat{u}$ and $\hat{l}$ correspond to  upper and lower bands respectively.  
We consider the scenario wherein the lower band is completely filled (henceforth it will be ignored) whereas the upper band is filled till the Fermi momentum $\pm k_F$. Thus the field operator $\psi(x)$ 
has the following form
\begin{align}
\psi(x)=\big[\hat{u}(k_F)\psi_R(x)e^{ik_Fx} +\hat{u}(-k_F)\psi_L(x)e^{-ik_Fx}\big],\nonumber 
\end{align}
where $\psi_R(x)$ and $\psi_L(x)$
are the slow degrees of freedom about the points $k_F$ and $-k_F$, respectively, and the fermion spin texture is given by 
\begin{align}\hat{u}(p) = \frac{1}{2}\big\{  a_{+}+\frac{p}{|p|} a_{-},a_{+}-\frac{p}{|p|} a_{-}\big\},\end{align}
where $a_{\pm}=\sqrt{1\pm\frac{B}{ \sqrt{B^2+p^2}}}$.

We express $\psi_R(x)$ and $\psi_L(x)$ in terms of the slowly varying  bosonic fields $\phi(x)$ and $\theta(x)$ as follows
\begin{align}\label{Ffields}
\psi_R(x)=\frac{1}{\sqrt{2\pi a_0}}e^{i(\theta-\phi)},~~\psi_L(x)=\frac{1}{\sqrt{2\pi a_0}}e^{i(\phi+\theta)},
\end{align}
where $a_0$ is the short distance cutoff and the bosonic fields satisfy the  commutation relation: $[\phi(x),\theta(y)]=-i\pi\text{sign}(x-y)/2 $.  Plugging~(\ref{Ffields}) in to~(\ref{HL}) the  Hamiltonian acquires the  standard quadratic  form in terms of the bosonic fields~\cite{giamarchi}
\begin{align}\label{Hfree}
H_{\text{HL}}^0=v_F\int \frac{dx}{2\pi} [(\partial_x \phi)^2 + (\partial_x \theta)^2].
\end{align}
The  Hamiltonian~(\ref{Hfree}) is modified by including the interaction terms $1/2\int dx dx' U_e(x-x')\rho(x)\rho(x') $, where the density operator is given by
\begin{eqnarray}
&&\rho(x) =\psi_R^{\dagger}(x)\psi_R(x) +\psi_L^{\dagger}(x)\psi_L(x)+\frac{B}{\sqrt{B^2+k_{F}^{2}}}{}\nonumber\\ && \times[\psi_R^{\dagger}(x)\psi_L(x)e^{-i2k_Fx} +\psi_L^{\dagger}(x)\psi_R(x)e^{i2k_Fx}].
\end{eqnarray}
It is worth noting that the $2k_F$ component of the density in a helical liquid is allowed due to the presence of the magnetic-field.
The interaction corrections 
arising from the  forward-scattering terms: 
$\psi_{R/L}^{\dagger}(x)\psi_{R/L}(x)\psi_{R/L}^{\dagger}(y)\psi_{R/L}(y)$ and $\psi_{R/L}^{\dagger}(x)\psi_{R/L}(x)\psi_{L/R}^{\dagger}(y)\psi_{L/R}(y)$ yield $\frac{\tilde{U}_e(0)}{2\pi^2} \int dx (\partial_x \phi)^2$ term to the Hamiltonian, where $\tilde{U}_e(k)$ is the $k^{th}$ mode of the $U_e$ potential. On the other hand,   from the 
back-scattering terms $$\psi_{R/L}^{\dagger}(x)\psi_{L/R}(x)\psi_{L/R}^{\dagger}(y)\psi_{R/L}(y)e^{\mp i2k_F(x-y)},$$ one obtains correction to the Hamiltonian  which is proportional to the square of the field-strength 
and given by~\cite{Zyuzin2010}
$$ -\frac{B^2}{B^2+k_F^2}\frac{\tilde{U}_e(2k_F)}{2\pi^2} \int dx (\partial_x \phi)^2.$$ 
The interaction modified Hamiltonian  thus acquires the following form
\begin{align}\label{HInt}
H_{\text{HL}}=v\int \frac{dx}{2\pi} [\frac{1}{K}(\partial_x \phi)^2 + K[\pi \Pi
(x)]^2],
\end{align}
where $\Pi(x)= \partial_x\theta(x)/\pi$,  $v=\sqrt{v_F(v_F+r)}$, $K=\sqrt{v_F/(v_F+r)}$ and $r=  [\tilde{U}_e(0)-B^2/(B^2+k_F^2)\tilde{U}_e(2k_F)]/\pi$. In terms of the  bosonic annihilation operator 
\begin{align}\label{ann}
b_p=\frac{1}{\sqrt{2|p|K}}  [-\frac{|p|\phi_p}{\sqrt{\pi}}  +  i\frac{K}{\sqrt{\pi}} \Pi_p],
\end{align}
the potential term due to the 	interaction of the mobile  impurity  with the bosonic excitation, $V=U\int dx a^\dagger (x)a(x) \rho(x)$,  is
given by 
\begin{eqnarray}\label{Eq:Int1D}
V&=& U'\sum_{k,q}i\text{sgn}(q)\sqrt{\frac{|q|}{2\pi  L}}a^{\dagger}_{k+q}a_k (b_q+b^\dagger_{-q}),
\end{eqnarray}
where  $U'=UK$. We have neglected the  large momentum transfer terms as we consider the simpler scenario for which the  $B-$field is switched off.\\

Thus the full hamiltonian with the impurity interaction term acquires the form 
$$	H=\sum_{k}\epsilon_k a^\dagger _k a_k +v \sum_p |p| b^\dagger_p b_p  +V.$$
With this expression for the Hamiltonian, we will employ the  linked cluster expansion technique to describe the modifications to the impurity Green's function~\cite{Kantian2014}. 
As before, the interaction modified  impurity Green's function has the form 	$ G(k,t)= G_0(k,t)e^{\sum_i \mathcal{S}_{i}}$, where $ G_0(k,t)=-i \theta(t)e^{-i\epsilon_p t}$. It suffices to focus till the second cummulant. The first cumulant,  $\mathcal{S}_1 =-i\int dt_1 \langle|a_k(t)V(t_1)a^\dagger _k (0)|\rangle/G_0$ vanishes as it involves averaging over a single boson
operator. The non-vanishing contribution arises from the second cumulant:
$\mathcal{S}_2(t)=G_0^{-1}\mathcal{M}
_2-\mathcal{S}_1^2/2$, where
$$\mathcal{M}_2(t)=\frac{(-i)^3}{2}\int dt_1 \int dt_2 \langle|a_k(t)V(t_1)V(t_2)a_k^\dagger (0)|\rangle.$$
As in the 2D case 
only the connected diagrams need be considered. 
In terms of the unperturbed Green's function the second cumulant has the following form,

\begin{eqnarray}
&&	\mathcal{S}_2(t)=(-i)^3\sum_{q}V^2(q)\int_0^t dt_1 \int_0^{t_1} dt_2 G_0(k,t-t_1)\notag \\
&& G_0(k+q,t_1-t_2)G_0(k                                                                                                                                                                                                                                                                                                                                                                                                                                                                                                                                                   ,t_2)D_0(q,t_1-t_2)/G_0(k,t),
\end{eqnarray}
where $D(q,t_1-t_2)=-i\theta(t_1-t_2)e^{-iv|q| (t_1-t_2)}-i\theta(t_2-t_1)e^{iv|q| (t_1-t_2)}$ is the zero temperature time ordered bosonic Green's function. Performing the integration  over $t_2$ and $t_1$ we obtain

\begin{eqnarray}\label{S2_1D}
\mathcal{S}_2(k,t)=-\int d\omega \rho(\omega,k)\bigg[-\frac{it }{\omega}+\frac{1 - e^{-it\omega }}{\omega^2}
\bigg],
\end{eqnarray}
where
\begin{eqnarray}\label{rho-1D}\rho(\omega,k)=\frac{U'^2}{2\pi}\int \frac{dq}{2\pi} |q|\delta(\omega-\epsilon_{k+q}+\epsilon_k-v|q|).
\end{eqnarray}
We note that similar to the 2D case, the first term (linear in time term) in ~(\ref{S2_1D}) renormalizes the impurity,   whereas it is again the second term which determines the long time asymptotics of the impurity Green's function.

\subsection{Non-recoil case}
For the non-recoil case which also corresponds to $M=\infty$,  the impurity energy terms drop out from the $\delta$-function, therefore the  $\rho$ 
term acquires the simple form
\begin{equation}
\rho(\omega)=\frac{U'^2}{2\pi }\int \frac{dq}{2\pi}~|q|\delta(\omega-v|q|)=\frac{U'^2}{2\pi^2 v^2 }\omega,
\end{equation}
where $\omega>0$. 
The long time asymptotics in particular the decay of impurity Green's function is determined by the  following term of $\mathcal{S}_2$
$$ -\frac{U'^2}{2\pi^2 v^2 }\int \frac{d\omega (1-e^{-i\omega t})}{\omega}\approx  -\frac{U'^2}{2\pi^2 v^2 } \log(t\omega_c).$$
The Green's function thus has a power-law decay given by
\begin{equation}\label{GF:decay1}G(t) \propto t^{-\frac{U'^2}{2\pi^2v^2}},
\end{equation}
resulting in a non-Lorentzian spectral function.
The above calculation confirms the well-known fact that in a 1D system the  introduction of heavy impurity  leads to orthogonality catastrophe.\\

\subsection{Recoil case}
Consider first the scenario	 for small  impurity momentum, in particular $k\ll Mv$. Unlike the 2D case, where the impurity exhibits quasiparticle behavior even at very low momenta, in 1D the decay-behavior of the Green's function  remains unchanged and is given by Eq.~\ref{GF:decay1} implying a non-quasiparticle behavior. 
Consider next the scenario $k< Mv$, but $(Mv-k)/k\sim 1$. The long-time behavior of the impurity  is determined by 
$\rho$  near the small frequencies and the corresponding $\omega$ expansion of   $\rho$  yields the following form 
\begin{eqnarray}
\rho(\omega)&=&\frac{U'^2M}{2\pi }\int \frac{dq}{2\pi}|q|\sum_{i=1}^{2}[\frac{\delta(q-\frac{M\omega}{Mv+(-1)^ik})}{Mv+(-1)^ik}]\notag\\
&=&\frac{U'^2M^2\omega}{2\pi^2} \Big[ \frac{M^2v^2+k^2}{(M^2v^2-k^2)^2}\Big].
\end{eqnarray}

The Green's function, therefore, exhibits power-law decay given by

\begin{eqnarray}
G(k,t) \propto t^{-\frac{U'^2}{2\pi^2}\frac{v^2+k^2/M^2}{(v^2-k^2/M^2)^2}},
\end{eqnarray}
where the exponent is now $k-$dependent and the  $k/Mv\ll 1$ limit (\ref{GF:decay1}) is recovered from the above equation. Inspite of the decay behavior, for $k\gg \sqrt{2M/\tau_0}$ (where $\tau_0= e^{2\pi^2v^2/U'^2}/\omega_c$) a  quasiparticle type behavior is expected till time $t\sim\tau_0$.

Finally consider the scenario wherein the initial impurity momentum is large, i.e., $k> M v$. 
In this case, the main contribution from the  $\delta$-function integration yields  a frequency independent term, $\rho(\omega)= U'^2 M/2\pi^2$, arising from  the $q \approx 2(k-Mv)$ region.  
Thus from  Eq.~(\ref{S2_1D}), it is easy to deduce that the decay term of the Green's function  results in  a conventional Fermi-liquid type term~\cite{Kantian2014}, i.e.,  $e^{-t/\tau}$ 
where  the life-time is  given by $1/\tau \approx U'^2M/4\pi $ (thus for $v\gg U'$ the excitation is well defined).  The oscillatory term on the other hand   acquires contribution from a rather unusual  term given by $(U'^2 M/2\pi^2)t\log t\omega_c$, which can be neglected in comparison  to  $k^2/2M$ for $t<\tau$ as long as $v\gg U'\sqrt{\log (\omega_c/U'^2M)}$.
This criterion on $v$ also implies that the subleading contribution from the second $q-$region ($\approx M\omega/(Mv+k)$ where the $\delta$-function is non-zero) can be neglected.\\

\section{Mobility of Impurity in 1D}{\label{sec:VI}}
The temperature dependence of the mobility of impurity constrained  to move in 1D and interacting with the helical edge modes exhibit contrasting behavior in the presence and the absence of a magnetic field.
We will again focus our attention on the low temperture regime $T\ll k_F^2/M$.
Consider first the  scenario without the  magnetic field, as discussed earlier the back-scattering processes will be  absent and 
only  the forward scattering  processes governed by the interaction term
~(\ref{Eq:Int1D}) are allowed. 
Focussing on the weak-coupling limit we utilize the Boltzmann equation approach to analyze the temperature depedence of the  mobility.  In the presence of an external  electric field $E$, the steady state Boltzmann equation for the momentum distribution function $f_{p,t}$ is given by	
\begin{eqnarray}\label{Eq:Boltz}
eE \frac{\partial f_{p,t}}{\partial p}=\sum_{k} [f_{k,t}\Gamma(k; p)-f_{p,t}\Gamma(p; k)],
\end{eqnarray}
where $e$ is taken to be the charge of the heavy particle. The effects of the scattering processes are encoded on the RHS which is also  the collision integral. As in the 2D case, the equilibirum distribution function of the impurity is given by the Maxwell-Boltzmann distribution function $f^0_k = N e^{-\beta k^2/2M }$, where the normalization constant is $N = \sqrt{ 2\pi \beta/M}$.  Indeed, in the equilibrium scenario the LHS vanishes, 
therefore, the following detailed balance equation $ f^0_{k}\Gamma(k; p)=f^0_{p}\Gamma(p; k)$ is neccessarily satisfied.
The scattering rate $\Gamma$  obtained using the Fermi-Golden rule has the form
\begin{eqnarray}
\Gamma(k; p) =\frac{U^2}{vL}\Big[\omega_q (n_q+1)\delta(\frac{p^2}{2M}-\frac{k^2}{2M} +\omega_q)+\notag\\
\omega_q n_q\delta(\frac{p^2}{2M}-\frac{k^2}{2M} -\omega_q)\Big],\quad
\end{eqnarray}
where  $q=p-k$ and $n_q$ is the equilibriun  Bosonic distribution function. 
Consider the first term of~(\ref{Eq:Boltz}) where the summation in $k$ implies that the typical impurity momentum is  $k\sim\sqrt{MT}$ while the energy conservation criterion forces phonons with momentum  $q\sim Mv\approx Mv_F$  to take part in the scattering process, however, this is an exponentially rare process since $T\ll Mv_F^2$. 
Thus the contribution to friction due to this term and following similar arguments due to the second term is exponentially suppressed. 
Consequently, mobility diverges exponentially. 

Turning on the magnetic field opens up the  back-scattering channel,  thus these processes can in principle yield finite contributions to the mobility. The   interaction term  now has an additional term given by 
\begin{eqnarray}
\label{eq:BS}
\frac{U}{2\pi a_0}\frac{B}{\sqrt{B^2+ k_F^2}}\int dx a^\dagger(x) a(x) \cos(2\phi-2k_F x). 
\end{eqnarray}
Consider the possibility of $2k_F$ momentum transfer to the impurity particle with  momentum $k\sim \sqrt{MT}$, in this case the energy transferred will be   $\sim k_F^2/M$. Since the temperature regime we are considering is much smaller than this energy scale,
it is again an exponentially suppressed process. 
Thus unsurprisingly this process will also not cause impediments to the impurity flow.\\

It turns out that even though the  second order process arising from~(\ref{eq:BS}) is perturbatively weaker in comparison to
the first order back-scattering process, yet it yields dominant contribution to the scattering rate at low temperatures.  The interaction term for a  second order process  can be written as~\cite{NetoF}
$$V_{2}=\mathcal{V}_2 \int dx a^\dagger (x) a(x)\psi^\dagger_R\psi_R\psi^\dagger_L\psi_L,$$
where $\mathcal{V}_2=U^2B^2/[(B^2+k_F^2)\epsilon_{2k_F}]$. In terms of the bosonic annihilation operator~(\ref{ann}) the interaction term is given by, 
\begin{eqnarray}\label{NetoV}
&&V_{2}=-\frac{\mathcal{V}_2}{8 \pi L}\sum_{p,k_1,k_2}\sqrt{|k_1| |k_2|}~a^\dagger_{p+k_1+k_2}a_{p}
\Big[\Big(\frac{k_1 k_2}{|k_1 k_2|} K- K^{-1}\Big){}\nonumber\\
&&\times b^\dagger_{-k_1}b^\dagger_{-k_2}+\Big(\frac{k_1 k_2}{|k_1 k_2|} K+ K^{-1}\Big) b_{k_1}b^\dagger_{-k_2}\Big]+h.c.
\end{eqnarray}

The first term of the above equation represents  a scattering process which involves a mobile impurity with  an initial momentum $p$ getting scattered into the state $p+k_1+k_2$ via the creation of two phonons with momentum $-k_1$ and $-k_2$.  This process requires the initial energy of the mobile impurity to be $\sim Mv_F^2$ and hence an unfavorable process.  
Similar argument holds for its hermitian conjugate pair. 
We can therefore approximate~(\ref{NetoV}) as 
\begin{eqnarray}\label{NetoV1}
V_{2}\approx-\frac{\mathcal{V}_2}{4 \pi L}\sum_{p,k_1,k_2}&&\sqrt{|k_1| |k_2|}
\Big(\frac{k_1 k_2}{|k_1 k_2|} K+ K^{-1}\Big){}\nonumber\\
&& a^\dagger_{p+k_1+k_2}a_{p}b_{k_1}b^\dagger_{-k_2}.
\end{eqnarray}
The interaction term now represents the scattering of the mobile impurity via the  destruction and creation of phonons. The requirement for this  scattering process to be relevant is that  both the initial and final energies of the mobile impurity and the phonons are   $\sim T$. As will be discussed below this requirement is satisfied.
The  collision integral i.e., the RHS of~(\ref{Eq:Boltz}) with this new interaction term is given by
\begin{eqnarray}\label{Eq:coll}
I(p)= \sum_{q} [-f_{p}\Gamma_2(p; p+q)+f_{p+q}\Gamma_2(p+q; p)],
\end{eqnarray}
where $\Gamma_2$ is the scattering rate.  Defining  the non-equilibrium distribution function as $f_{p}=f^0_{p} h_p$ and using a similar  detailed balance equation as discussed earlier  one obtains,
$I(p)= \sum_{q} f^0_{p+q} (h_{p+q}-h_{p}) \Gamma_2(p+q; p).$  Using Fermi's golden rule the full expression for the  collision integral can be written as
\begin{widetext}
\begin{eqnarray}\label{Eq:coll2}
I(p)= \frac{\mathcal{V}_2^2}{32 \pi^3 }
\int dq d\bar{q}  
|k_1k_2|(h_{p+q}-h_{p})
\Big(K^2+ K^{-2}+2\frac{k_1 k_2}{|k_1 k_2|} \Big) f^0_{p+q}n_{k_1}(n_{k_2}+1)	\delta\Big(\frac{p^2}{2M}-\frac{(p+q)^2}{2M}+\omega_{k_2}
-\omega_{k_1}\Big),
\end{eqnarray}
\end{widetext}
where  $k_2= -(q+\bar{q})/2$ and  $k_1= (\bar{q}-q)/2$.

\begin{figure}
\centering
\includegraphics[width=1.0\linewidth]{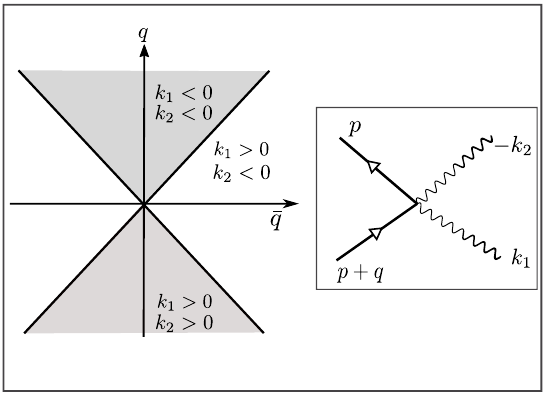}
\caption{(Color online) The scattering process can be divided into two regions shown by the shaded and the unshaded region in the $(q,\bar{q})$ plane.  The solid lines represent the scattering of the impurity and the wiggly  lines represent the phonons.}
\label{FEY1}
\end{figure}

\begin{figure}
	\centering
	\includegraphics[width=1.0\linewidth]{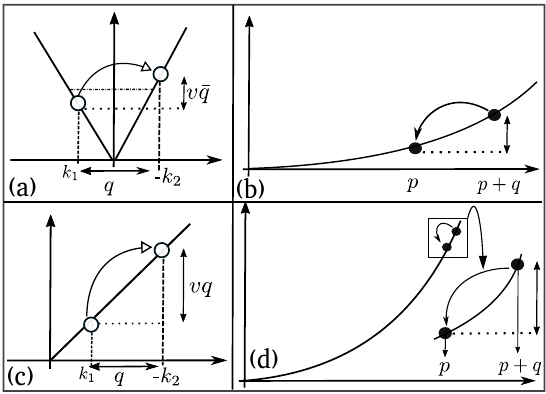}
	\caption{(Color online)~(a)~Represents phonon scattering in the $q>0$ and $q>\bar{q}$ regions. The momentum transferred is $q$, however, the energy transfer is $v\bar{q}$.~(b) The energy change of the mobile impurity: $p^2/2M-(p+q)^2/2M$. (c) ~Represents phonon scattering in the $\bar{q}>0$ and $\bar{q}>q$ regions. The momentum transferred is $q$, and  the energy transferred is $vq$.(d) The energy change of the mobile impurity has formally the same expression $p^2/2M-(p+q)^2/2M$. However, the energy-momentum constraint is satisfied for $q\sim Mv_F$ and with the corresponding energy of the phonon $\sim Mv_F^2$. An alternate description is given in the main text.}
	\label{ENG}
\end{figure}

The  evaluation  of the $\delta$-function can be divided into the  following two cases, $\omega_{\frac{|q+\bar{q}|}{2}}-\omega_{\frac{|\bar{q}-q|}{2}} =\pm v q$   and $\omega_{\frac{|q+\bar{q}|}{2}}-\omega_{\frac{|\bar{q}-q|}{2}}=\pm v \bar{q}$. These   are  the unshaded and the shaded regions of Fig.~\ref{FEY1}, respectively. The former  scenario is irrelevant since the $\delta$-function imposes     constraint similar to the one discussed before, i.e., the requirement that the phonons have energy $\sim Mv_F^2$.  The later case on the other hand is achieved for the range  $q^2 > \bar{q}^2$ where the momentum transfer $ q$ changes the direction of phonons, i.e., $k_1$ and $-k_2$ are in the opposite direction, however, the energy of phonon hardly changes (see~Fig.\ref{ENG}).   This is reflected from  the $\delta$-function constraint which fixes the energy transfer to    $|v\bar{q}|=|\xi_{p+q}-\xi_{p}|$, where $|\bar{q}/q|\sim\sqrt{T/Mv_F^2} \ll 1$. Therefore~(\ref{Eq:coll2}) reduces to

\begin{eqnarray}\label{Eq:coll3}
I(p)= \frac{\mathcal{V}_2^2  (K+ K^{-1})^{2}}{128 \pi^3 }
\int dq &&q^2 f^0_{p+q}(h_{p+q}-h_{p})\nonumber\\
&& \times 	n_{\frac{q}{2}}(n_{\frac{q}{2}}+1).
\end{eqnarray}

We will next consider the limit of weak electric field $E$. Following Feynman~\emph{ et al.,}~\cite{Feynman}  $h_p$ is expanded  to linear order in $E$
as $h_p=1+ pE{\mathcal{H}}$, where $\mathcal{H}$ is a weakly varying  even function of $p$. 
The integral is evaluated to yield
\begin{align}\label{Eq:coll4}
I(p)=\frac{2\pi T^5}{15}\mathcal{V}_2^2 (K+ K^{-1})^{2} E{\mathcal{H}}\frac{\partial f^0_p}{\partial p}.
\end{align}
Under the steady-state condition, the LHS of~(\ref{Eq:Boltz}) is simply given by $eE\partial f_p \approx eE\partial f^0_p (1+ pE{\mathcal{H}})$. Thus comparing it with~(\ref{Eq:coll4}), we obtain 
$$\mathcal{H} \approx \frac{15}{2\pi T^5} \frac{e }{\mathcal{V}_2^2(K+ K^{-1})^{2}}.  $$
With the non-equilibrium distribution determined, the mobility, $\mu$,  of impurity in the presence of electric-field  $E$ can be easily calculated and is 
given by 
\begin{align}
\mu = \frac{\int dp p^2 f_0(p)  E{\mathcal{H}}/M}{E \int dp f_0(p)}=  \mathcal{H} T\propto \frac{1}{T^4B^4}.
\end{align}
Thus with the help of the  above ansatz it is easy to deduce that the mobility  diverges as $T^{-4}$. Similar conclusion 
was reached in an earlier work~\cite{NetoF} using 
an alternate approach.
It is worth noting that inhere the power-law divergent   behavior is achieved only in the presence of a magnetic field. In the absence of a magnetic field the mobility diverges exponentially at low temperatures.

\section{Summary}

To summarize, we have presented a detailed study of the Green's function and the mobility of a single nonmagnetic impurity interacting with the bath of 2D  and 1D Dirac fermions.  

In the 2D scenario, impurity  Green's function exhibits different behavior in the non-recoil and the recoil case.
A crucial ingredient for the analysis is the density of particle-hole excitations evaluated by the momentum integration of the imaginary part of the polarization function of the Dirac fermions.
The non-recoil case results in the generation of a large number of particle-hole excitations whose density varies linearly with 
$\omega$ and this, in turn, results in a power-law decay of the  Green's function $\propto 1/t^\nu$, where $\nu =k_F^2U^2/\pi^2$.
The impurity can no longer be described in terms of the quasiparticle picture, in particular, the spectral-function is modified from  $\delta$-function and manifests a sharp cut-off for energies less than the renormalized impurity energy,  while a power-law suppression is exhibited for energies greater than it, given by $A(\epsilon)\propto \Theta (\epsilon - \epsilon_p)/(\epsilon - \epsilon_p)^{1-\nu}$. In contrast, the energy-momentum constraint in the recoil case implies reduced phase-space for the particle-hole excitations resulting in a $\omega^{3/2}$ dependence of the density of states. The resulting Green's function has a pure oscillatory part, implying non-zero quasiparticle weight, in addition, an oscillatory part multiplied by a decaying $t^{-1/2}$ term. While the former is responsible for a delta-function peak in the spectral function, the latter yields an incoherent part that exhibits square-root singularity.

The temperature dependence of the mobility of the impurity has been estimated by performing a statistical average on the inverse quasiparticle lifetime with respect to the Boltzmann weight factor.  
The mobile impurity interacts with a particle-hole excitation having typical energy $\omega\sim T$ and momentum $q\sim \sqrt{2MT}$. In this regime, the polarization function acquires a particularly simple form and the temperature dependence of the mobility is revealed to be $T^{-3/2}$. 

For the case of a mobile impurity interacting with the 1D helical modes, 
similar to the Green's function behavior in 2D, the Green's function in the non-recoil case exhibits power-law suppression at long times with $G\sim t^\nu$, where $\nu=-U'^2/2\pi^2v^2$. Unlike the 2D case, this behavior persists even for the recoil scenario,   albeit for a finite range of momentum. In particular, for $k< Mv$, the long-time decay exponent of the Green's function acquires a momentum dependence  in the exponent given by $$\nu =-\frac{U'^2}{2\pi^2}\frac{v^2+k^2/M^2}{(v^2-k^2/M^2)^2},$$ whereas for $k>Mv$
the Green's function has a conventional Fermi-liquid type of decay with the decay time given by $\tau^{-1}=U'^2M/4\pi$.

The temperature dependence of the mobile impurity interacting with the 1D helical modes exhibits contrasting behavior with or without the magnetic field. In the absence of a magnetic field only the forward scattering process is allowed, the energy and momentum constraint forces exponential divergence of the mobility as the temperature is lowered.  
Turning on the magnetic field opens up the back-scattering channel, nevertheless, at the lowest order in interaction, the mobility retains the exponential divergence. However,
the second-order back-scattering process allows a scattering process in which the energy transferred between the mobile impurity and the phonons is negligible compared to the temperature. Using the 
Feynman's ansatz we solve the Boltzmann equation to obtain $T^{-4}$ divergence of the mobility which also diverges with respect to the magnetic field as $B^{-4}$.

\section{ACKNOWLEDGMENTS}
We would like to thank B. Braunecker and V. Zyuzin for useful discussions. S.G. is grateful
to SERB for the support via the grant number
EMR/2016/002646.

\section{Appendix}\label{pol_func}

The noninteracting generalized polarization function for the Dirac fermion in the TI is given by
\begin{eqnarray}
\Pi(q,\omega_n)=-\int _{\text{K}} \text{Tr}\Big[\sigma_0\mathcal{G}_{\text{K}}\sigma_0\mathcal{G}_{{\text{K+Q}}}\Big],\label{eq:PF}
\end{eqnarray}  
where  $\text{Tr}$ denotes the trace, ${\text{K}}=(\vec{k},\Omega)$ and  ${\text{Q}}=(\vec{q},\omega)$. 
The corresponding zero temperature single particle Matsubara Green's function used in the above equation has the following form
 \begin{equation}
\mathcal{G}(k,i\Omega)=\frac{1}{2}\sum_{\alpha=\pm 1}\left[\frac{\hat{I}-\alpha(\vec{\sigma}\cdot \vec{\bar{k}})/\xi_k}{i\Omega_n+\alpha\,\xi_k+\mu_F}\right],
\end{equation}
where $\alpha= \pm 1$ represents valence and conduction bands respectively,  $\vec{\bar{k}}=k_x\hat{e}_1+k_y\hat{e}_2+\Delta\hat{e}_3$,
and $\xi_k= \sqrt{k^2+\Delta^2}$.
The Pauli matrix $\sigma$ acts on the spin degrees of freedom.
Following the standard frequency summation and  the analytical continuation $i\omega\rightarrow \omega+i0^+$, we obtain the following form of the polarization function,
\begin{eqnarray}
\Pi(q,\omega) = -\int \frac{d^2k}{(2\pi)^2}\sum_{\alpha,\alpha'=\pm1}\Bigg[1+ \alpha \alpha'\frac{\vec{k}_\cdot\big(\vec{k}+\vec{q} \big)}{\xi_k\xi_{k+q}} \Bigg]\notag\\
\times \frac{n_F(-\alpha \xi_k)-n_F(-\alpha'\xi_{k+q})}
{\Big(\alpha\xi_k-\alpha'\xi_{k+q}-\omega - i0^+\Big)}.\qquad
\end{eqnarray}

The nonzero contribution to the  imaginary part of the polarization function from the upper to upper band $(u\rightarrow u)$ transitions is as follows,
\begin{widetext}	
\begin{eqnarray}
\text{Im}\Pi(q,\omega)=-\pi\int \frac{d^2k}{(2\pi)^2}\Bigg[1+\frac{\vec{k}_\cdot\big(\vec{k}+\vec{q}\big)}{\xi_k\xi_{k+q}}\Bigg]\Big(n_F(\xi_k)-n_F(\xi_{k+q})\Big)\delta(\omega+\xi_k-\xi_{k+q})
\end{eqnarray}
After delta-function intergration we obtain
\begin{eqnarray}
\text{Im}\Pi(q,\omega)=-\frac{1}{2\pi}\int_{{\text{Max}}[(\mu_F-\omega),\Delta]}^{\mu_F} \frac{d\xi_k}{\sqrt{q^2-\omega^2}}\Bigg[\frac{(2\xi_k+\omega)^2-q^2}{\sqrt{(2\xi_k+\omega)^2-\zeta^2}}\Bigg]
\end{eqnarray}
\[
\text{Im}\Pi(q,\omega) =- \frac{1}{2\pi\sqrt{q^2-\omega^2}}\,\,\,\times \left \{
\begin{tabular}{ccc}
$\mathcal{F}\big(2\mu+\omega\big)-\mathcal{F}\big(2\text{max}[\mu-\omega,\Delta]+\omega\big) \hspace{1.0cm}:1A$ \\
$\mathcal{F}\big(2\mu+\omega\big)-\mathcal{F}\big(\zeta\big) \hspace{0.5cm} \hspace{3.35cm}:2A$
\end{tabular}
\right \},
\]
\begin{eqnarray}
{\text{where}}\,\,\zeta=\sqrt{q^2 +4q^2\Delta^2/(q^2-\omega^2)},\quad{\text{and}}\quad \mathcal{F}(x)=\frac{1}{4}\Bigg\{\Big[\zeta^2-2q^2\Big]\log\big(\sqrt{x^2-\zeta^2}+x\big) +x\sqrt{x^2-\zeta^2}\Bigg\}.\label{App_ImPi}
\end{eqnarray}	
	
The allowed regions for the transitions are
\begin{eqnarray}
&&1A:\omega<\mu-\sqrt{(q-k_F)^2+\Delta^2}\notag\\
&&2A:\pm\mu\mp \sqrt{(q-k_F)^2+\Delta^2}<\omega<-\mu+\sqrt{(q+k_F)^2+\Delta^2}.\hspace{4.95cm}\notag
\end{eqnarray}
	
Next the simillar contribution from lower to upper band $(l\rightarrow u)$ transitions are,
\begin{eqnarray}
\text{Im}\Pi(q,\omega)=-\frac{1}{2\pi}\int_{\Delta}^{\omega-\mu_F} \frac{d\xi_k}{\sqrt{\omega^2-q^2}}\Bigg[\frac{-(2\xi_k-\omega)^2+q^2}{\sqrt{-(2\xi_k+\omega)^2+\zeta^2}}\Bigg]
\end{eqnarray}
	
\[
\text{Im}\Pi(q,\omega)=- \frac{1}{2\pi\sqrt{\omega^2-q^2}}\,\,\,\times \left \{
\begin{tabular}{ccc}
$\mathcal{F}'\big(\omega-2 \mu\big)-\mathcal{F}'\big(-\zeta\big) \hspace{1.8cm}:1B$ \\
$\mathcal{F}'\big(\zeta\big)-\mathcal{F}'\big(-\zeta\big) \hspace{2.75cm}:2B$ \\
$\mathcal{F}'\big(\zeta\big)-\mathcal{F}'\big(-\zeta\big) \hspace{2.75cm}:3B$\notag
\end{tabular}
\right \},
\]
\begin{eqnarray}
{\text{where}}\qquad \mathcal{F}'(x)=\frac{1}{4}\left[\big(2q^2-\zeta^2\big)  \tan^{-1}\bigg(\frac{x}{\sqrt{\zeta^2-x^2}}\bigg) +x\sqrt{x^2-\zeta^2}\right],
\end{eqnarray}
	
simillarly the allowed regions in the  $(q,\omega)$ plane are 
\begin{eqnarray}
&&1B :  \mu+\sqrt{(q-k_F)^2+\Delta^2}<\omega<\mu+\sqrt{(q+k_F)^2+\Delta^2}\notag\\
&&2B:\omega>\mu+\sqrt{(q+k_F)^2+\Delta^2}\notag\\
&&3B :  \omega>(2k_{F});\,\, \& \,\,\sqrt{q^2+4\Delta^2} <\omega< \mu+\sqrt{(q-k_F)^2+\Delta^2}.\notag
\end{eqnarray}
\end{widetext}

\bibliography{mobility} 
\end{document}